\documentclass[12pt]{iopart}

\usepackage{graphicx}
\usepackage{subfigure}

\begin{document}

\title[X-ray spectra in x-ray rooms]{X-ray spectra after shieldings in x-ray rooms}

\author{Antonio Gonz\'alez-L\'opez}

\address{Hospital Clinico Universitario Virgen de la Arrixaca - IMIB, ctra. Madrid-Cartagena s/n, El Palmar (Murcia), 30120, Spain}
\ead{antonio.gonzalez7@carm.es}
\vspace{10pt}
\begin{indented}
\item[]\today
\end{indented}

\begin{abstract}
Background: Determining the energy spectra of X-ray beams reaching a point after a protective barrier is necessary for the accurate calculation of absorbed dose at that point. The task is complex, given the diversity of both, the energy spectra reaching the barrier and the materials used for its construction.
Methods: This work provides spatial distributions of energy spectra of emergent photons after shielding barriers. These distributions are obtained through Monte Carlo simulations using exclusively monoenergetic pencil photon beams impinging perpendicularly on shielding barriers of different materials and thicknesses. The obtained results are applied to calculate the spectra of transmitted radiation in cases of polyenergetic broad beams similar to those encountered in clinical practice. Additionally, the calculated spectra are used to compute the dose transmitted through concrete barriers of various thicknesses for polyenergetic beams of 70, 100, and 140 kVp and compare the results with data from previous works.
Results: The importance of secondary radiation produced at the barrier through scattering or fluorescence of the incident photons in the total transmitted radiation varies drastically depending on the energy of the incident photons, the barrier material, and its thickness. The extension to polyenergetic beams demonstrates the high dependence of shielding efficiency on the spectrum energy of the scattered beam. Finally, the transmitted dose values calculated in this work are consistent with those found in the literature. Conclusion: The presented method, based on calculations of monoenergetic pencil beams, allows for the determination of absorbed dose in air in a large number of practical clinical situations, both outside conventional rooms and inside interventional rooms.
\end{abstract}

%
%
%
%
%

%
\section{Introduction}
To ensure radiological protection of individuals both outside and inside x-ray rooms, protective barriers are employed, which act as a shield between radiation sources and potentially exposed individuals. In some cases, the construction materials themselves (such as concrete, clay bricks, gypsum wallboard, or glass, among others) are sufficient to ensure an adequate level of protection. However, in other instances, additional shielding material (typically lead) is necessary to achieve that objective.

Several studies have described methods to properly design the necessary protections \cite{Archer1983,Simpkin1987,Petrantonaki1999,Tsalafoutas2003}, and some of them have been incorporated into international and national recommendations \cite{NCRP147}, enabling the design of nearly any medical x-ray installation today. 
At all times, new methods are being developed to facilitate the design of shielding for various facilities, such as dental radiology rooms \cite{Vega-Carrillo2022}, mammography suites \cite{Sampaio2013}, and those allowing for a better description of the protective materials used in radiology \cite{Eder2021,Edwards2016,Matyagin2016} and interventional procedures \cite{McCaffrey2012,Struelens2011}.

The factors determining the necessary thickness of the barriers to be employed are numerous. These include the energy of the radiation beam reaching the barrier, the material composition of the barrier itself, as well as factors such as the workload, the occupancy and use of the area to be protected and its classification.

The parameter typically used to characterize the energy of the radiation beam is its energy spectrum. However, the peak kilovoltage and beam filtration characterizinig the penetrability of the radiation, influence the shielding requirements and becomes one of the inputs in the tables used to calculate the beam transmission through a barrier \cite{Archer1983,Simpkin1987,Petrantonaki1999,Tsalafoutas2003,NCRP147,Vega-Carrillo2022,Sampaio2013,Eder2021,Edwards2016,
Matyagin2016,McCaffrey2012,Struelens2011}. In this way, calculating the attenuation for a range of clinically relevant beams can be carried out. However, the exact calculation of attenuation for polyenergetic beams with spectra different from those specified in standards, characterized by kilovoltage and filtration, is not possible. Such spectra often occur in practice, such as scattered radiation reaching the walls of the room or professionals within it. Nevertheless, if the energy spectra of transmitted radiation through a barrier are known for incident monoenergetic beams, these calculations are possible. The components of a polyenergetic beam incident on the barrier can be treated separately and added to obtain the polyenergetic beam emerging from the barrier. 

Kharrati et al. \cite{Kharrati2007,Kharrati2012} used Monte Carlo N-Particle (MCNP) \cite{mcnp} code to simulate monoenergetic broad beams to obtain x-ray buidup factors of lead, concrete and other building materials, and used them to calculate transmission factors for clinical beams of different kVp's and shielding barriers of different thicknesses. They compared their results with those obtained by Simpkin \cite{simpkin1989} who used Electron Gamma Shower (EGS4) \cite{egs4}.

Furthermore, if pencil beams are used, the principle of superposition ensures that the results obtained in terms of radiation spectrum at the barrier's exit will be valid for broad beams provided lateral equilibrium conditions are met \cite{simpkin1989,kato,GonzalezLopez2024}.

In this work, monoenergetic pencil beams are used to obtain the spatial distribution of energy of the photons emerging after barriers of different materials and thicknesses. Furthermore, the obtained results are applied to approximately describe practical cases encountered in the clinic: spectra behind a lead apron in an interventional radiology room are determined and spectra in the upper and lower floors of a conventional x-ray room are calculated. In addition, the calculated spectra are used to determine the dose transmitted through concrete barriers of various thicknesses for different polyenergetic beams.

\section{Material and methods}
\subsection{Monte Carlo simulations}
To describe the transmission of radiation energy through protective barriers, monoenergetic pencil beams incident perpendicular to the barrier were simulated. The barrier was modeled as an infinitely large plate with thickness $t$.

The simulated shielding materials were lead, concrete (portland), clay, gypsum (plaster of Paris), and glass. All of them were characterized by the materials defined in the PENELOPE \cite{PENELOPE} code, except for clay, which was defined using data from the compendium published by OSTI \cite{osti_1782721} in 2021 (material number 82). The simulated thicknesses were 0.1, 0.3, 0.5, 1.0, 1.5, 2.0, 3.0, and 5.0 mm for lead; 1, 5, 10 and 15 cm for concrete; 2, 5 and 15 cm for clay; 1 and 5 cm for gypsum; and 0.3 and 1.0 cm for glass.

The monoenergetic pencil beam transmission for these shielding materials was calculated for photon energies from 10 keV to 150 keV at 5 keV intervals for all the materials and thicknesses. Additional simulations were carried out for thicknesses 0.1, 0.5 and 1.5 mm of lead and 1, 5, 10 and 15 cm of concrete to complete incident energies from 10 keV to 150 keV at 1 keV intervals.

The Monte Carlo code used was PENELOPE\cite{PENELOPE}, and for each simulation, the number of histories run varied between $10^7$ and $10^8$. The absorption energies for electrons and photons were adjusted to $10^5$ and $10^3$ eV, respectively and transport parameters $C_1$ and $C_2$ were set to 0.1. These two constants determine the accuracy of electron transport by modeling the generation of random tracks, and using small values as 0.1 are recommended \cite{PENELOPE} for the energies studied in this work

\subsection{Polyenergetic beams in practical situations}
In this work, distributions of photons emerging after shielding barriers $N_{E_{in}}(E,r)$ at their exit surfaces are obtained by Monte Carlo calculations. In every case, the incident radiation is monoenergetic of energy $E_{in}$ and has an intensity of 1 photon, $E$ is the energy of the emerging photon and $r$ is its distance to the straight line defining the trajectory of the incident photon. For this kind of monoenergetic beams, the energy spectrum after the barrier can be obtained by
\begin{equation}
	N_{E_{in}}(E)=\int N_{E_{in}}(E,r)dr.
\end{equation}

Also, for an incident polienergetic beam with a spectrum $N_{in}(E_{in})$, the emerging beam will have the spectrum
\begin{equation}\label{eq:poly}
N(E)=\int N_{in}(E_{in})N_{E_{in}}(E)dE_{in}.
\end{equation}

The procedure for calculating spectra at the barrier's exit can be fed with any type of input spectrum found within the diagnostic radiology range. In particular, spectra from beams calculated in situations similar to those encountered in clinical practice can be used \cite{GonzalezLopez2022}. For some of these cases, the radiation spectra after the barrier are analyzed, including representative cases from interventional radiology and conventional radiology. For the former cases, a 0.3 mm lead protection barrier located 1 meter from a 16 cm diameter spherical water phantom is taken as the barrier. For the latter cases, the emergent radiation after 15 cm of concrete in the back-scatter and forward-scatter directions of a vertical beam is analyzed when the primary beam is directed towards a 32 cm diameter spherical water phantom phantom located at 3 m from the barrier. In these simulations, the absorption in the table and tube of the forward and back-scattered beams, respectively, has been ignored.

Spectra for primary poly-energetic x-ray beams, used to simulate those more clinically realistic primary beams, were calculated with Spektr 3.0 \cite{Punnoose2016}. 

Finally, the results obtained for the transmitted radiation spectra were used to calculate dose attenuation of the different simulated concrete barriers. To convert the fluence spectrum $\Phi(E)$ into dose, dose was approximated by the collision kerma $K_c$,
\begin{equation}\label{eq:kerma}
K_c=\int \Phi(E)E \frac{\mu_{en}(E)}{\rho}dE,
\end{equation}
where $\frac{\mu_{en}(E)}{\rho}$ is the mass energy absorption coefficient for air \cite{hubbel}. The dose transmission was then calculated as the ratio between the doses of the transmitted beam to the incident beam. 

The calculated attenuations were compared with those obtained in previous studies for barriers of varying concrete thicknesses. The beam spectra used correspond to the beams of constant potential employed by Kharrati et al. \cite{Kharrati2007} and Simpkin \cite{simpkin1989}. To achieve the HVLs that characterize these spectra (2.64, 3.77, and 5.26 mm Al), the filtrations of the 70, 100, and 140 kVp beams were adjusted to 2.024, 2.129, and 2.580 mm Al, respectively.

\section{Results}
\begin{figure}
\centering
\subfigure[0.5 mm lead, $E_{in}=60 \:keV$]
{\includegraphics[width=0.49\textwidth]{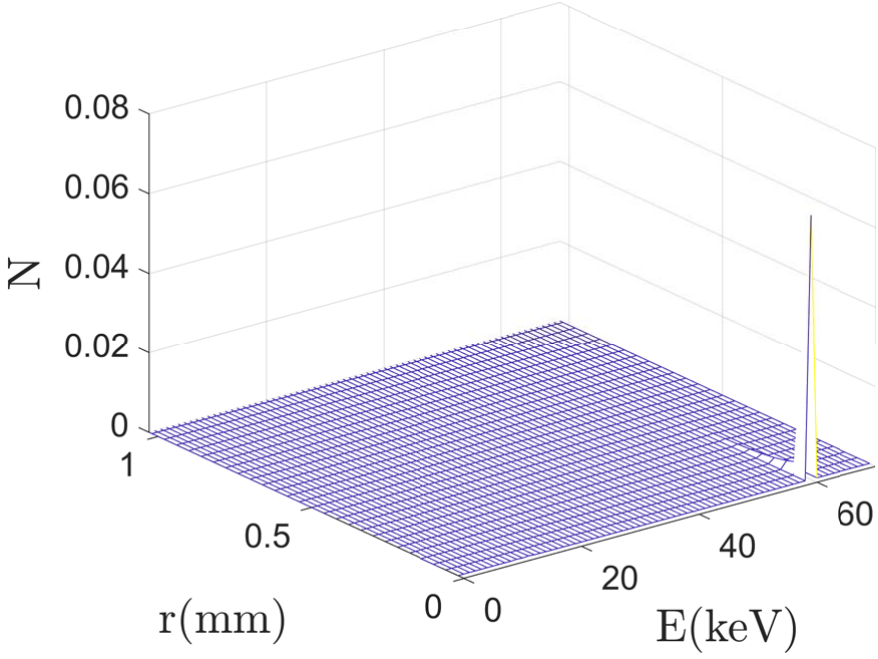}\label{fig:t_0p05cmLead60kVp}}
\subfigure[5 cm concrete, $E_{in}=60 \:keV$]
{\includegraphics[width=0.49\textwidth]{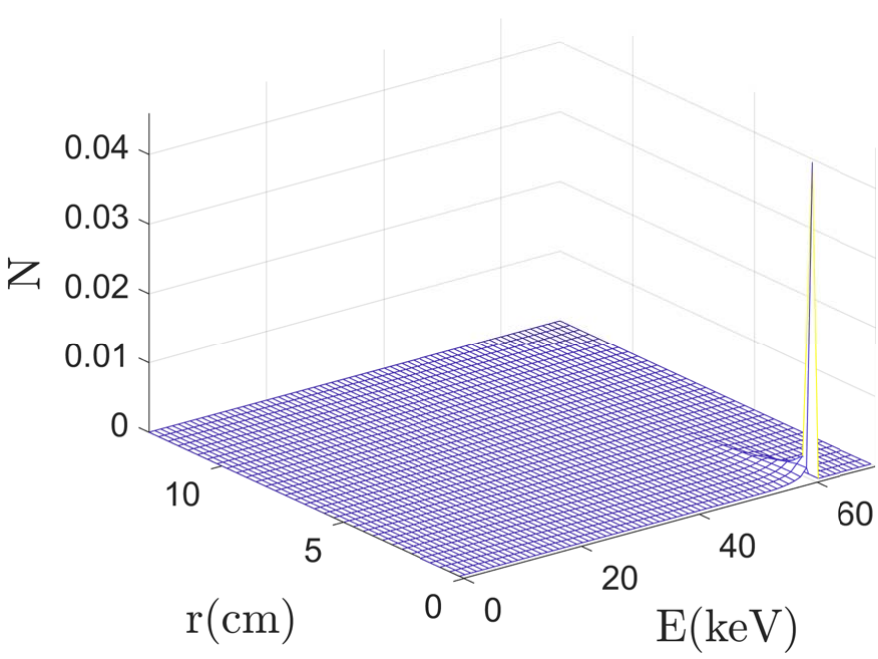}\label{fig:t_5cmConcrete60kVp}}
\subfigure[1.5 mm lead, $E_{in}=89 \:keV$]
{\includegraphics[width=0.49\textwidth]{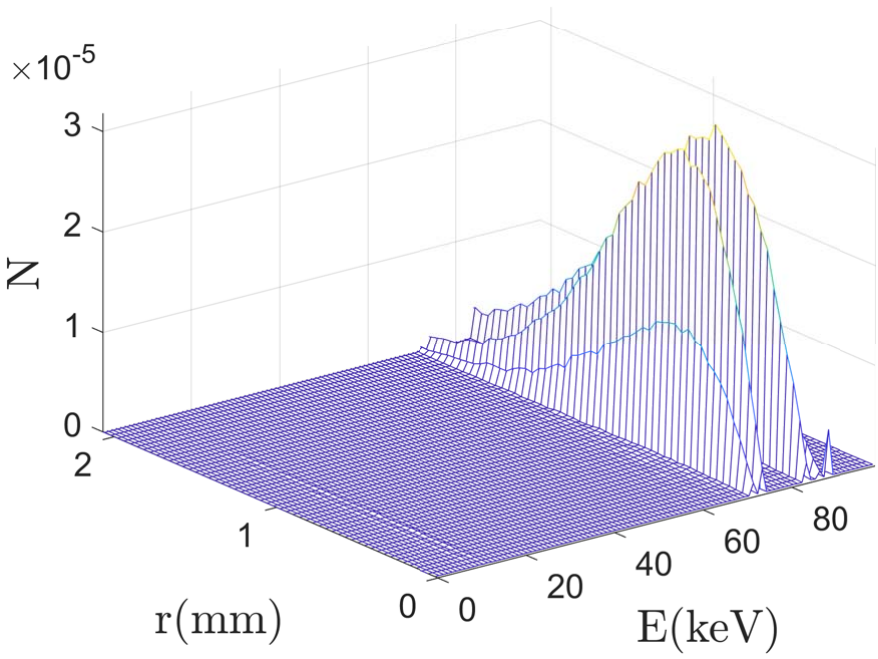}\label{fig:t_0p15cmLead89kVp}}
\subfigure[15 cm concrete, $E_{in}=89 \:keV$]
{\includegraphics[width=0.49\textwidth]{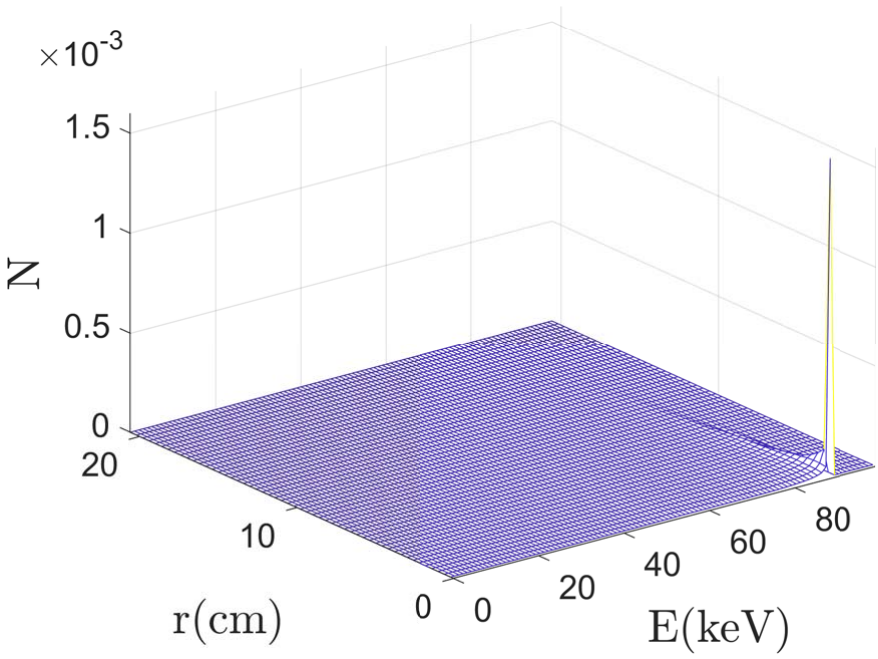}\label{fig:t_15cmConcrete89kVp}}
\subfigure[1.5 mm lead, $E_{in}=120 \:keV$]
{\includegraphics[width=0.49\textwidth]{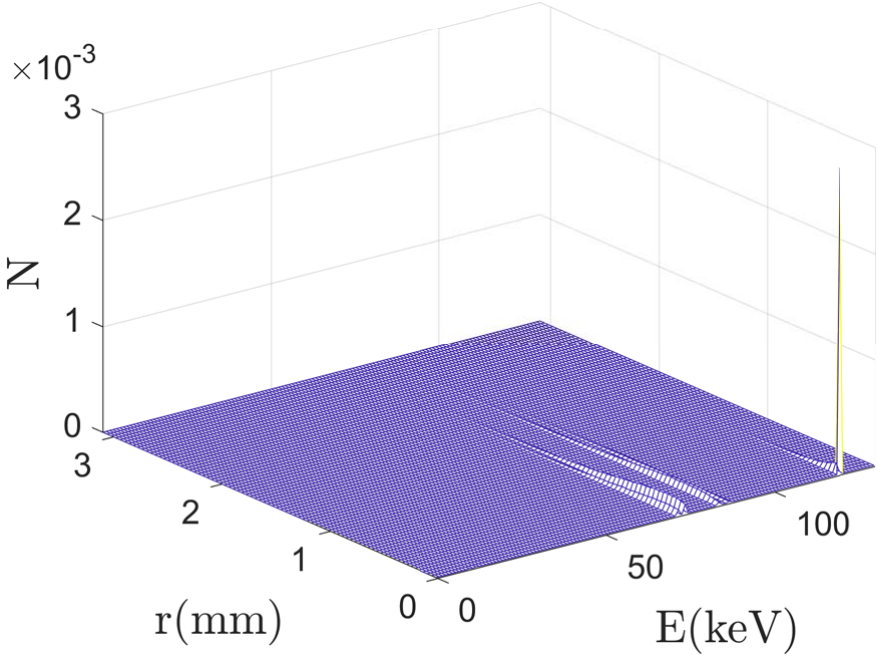}\label{fig:t_1p05cmLead120kVp}}
\subfigure[15 cm concrete, $E_{in}=120 \:keV$]
{\includegraphics[width=0.49\textwidth]{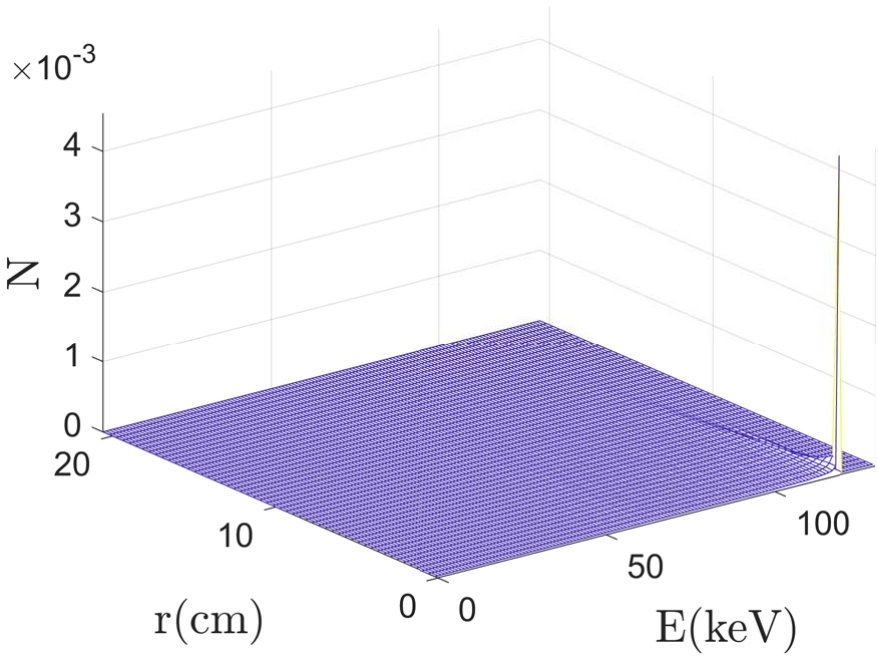}\label{fig:t_15cmConcrete120kVp}}
\caption{Distribution of transmitted photons per incident primary photon after shieldings of different materials and thicknesses. Three primary mono-energetic input pencil beams of energies 60, 89 and 120 keV are used and examples for lead thicknesses of 0.5 and 1.5 mm and concrete thicknesses of 5 and 15 cm are shown. Radial bins have a length of 0.04 mm for lead and 4 mm for concrete and energy bins have a length of 1 keV.}
\label{fig:spectra}
\end{figure}

Figure \ref{fig:spectra} shows the distribution of photons emerging after the barrier according to their radial position $r$ and energy $E$ per incident photon or energy $E_{in}$. Three monoenergetic pencil beams are used with energies $E_{in}$ of 60, 89 and 120 keV. The most striking aspect shown in the figure is the enormous modification that the spectrum of a beam with energy immediately above the K-edge of lead undergoes when it passes through a barrier of this material (figure \ref{fig:t_0p15cmLead89kVp}). 

\begin{figure}
\centering
\subfigure[Thickness: 1 cm]
{\includegraphics[width=0.49\textwidth]{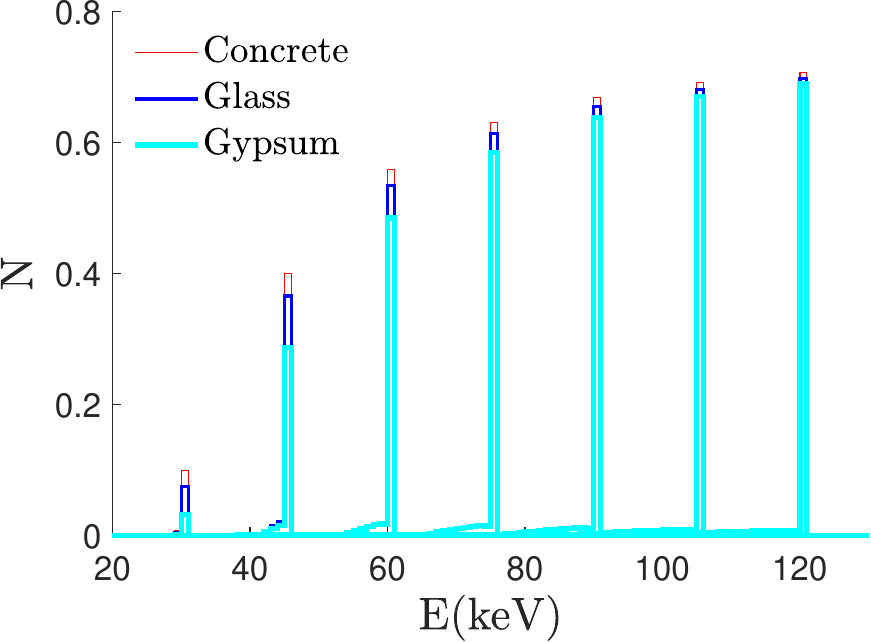}\label{fig:t1cm_horm_vidr_yeso}}
\subfigure[Thickness: 5 cm]
{\includegraphics[width=0.49\textwidth]{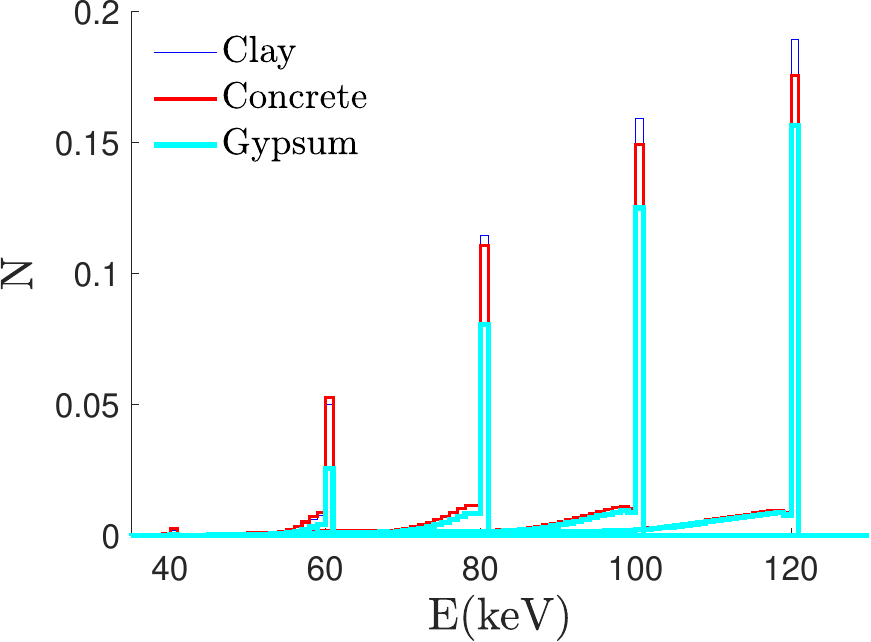}\label{fig:t5cm_horm_arc_yeso}}
\caption{Output spectra for monoenergetic incident beams as a function of the incident beam energy. Results are shown for two barrier thicknesses (1 and 5 cm) and four materials (concrete, glass, clay, and gypsum plaster of Paris\cite{PENELOPE}).}
\label{fig:materiales}
\end{figure}

Figure \ref{fig:materiales} shows the spectra of radiation emerging from barriers made of concrete, glass, clay, and gypsum, with results presented for thicknesses of 1 cm and 5 cm as a function of the incident photon energy (30, 45, 60, 75, 90, 105, and 120 keV for a 1 cm thickness and 40, 60, 80, 100, and 120 keV for a 5 cm thickness). In all cases, the behavior of the materials is similar, as they do not contain elements that exhibit K-shell fluorescence within the presented energy range. On the other hand, the large proportion of calcium in gypsum increases the material's effective atomic number compared to concrete, making it more absorbent to radiation. It should be noted that the simulated gypsum properties correspond to those of plaster of Paris powder (density 2.32 $g/cm^3$)\cite{PENELOPE}, not the building material gypsum wallboard which exhibits a large range of densities in practice \cite{osti_1782721}.

\begin{figure}
\centering
\subfigure[LI at 1 m]
{\includegraphics[width=0.49\textwidth]{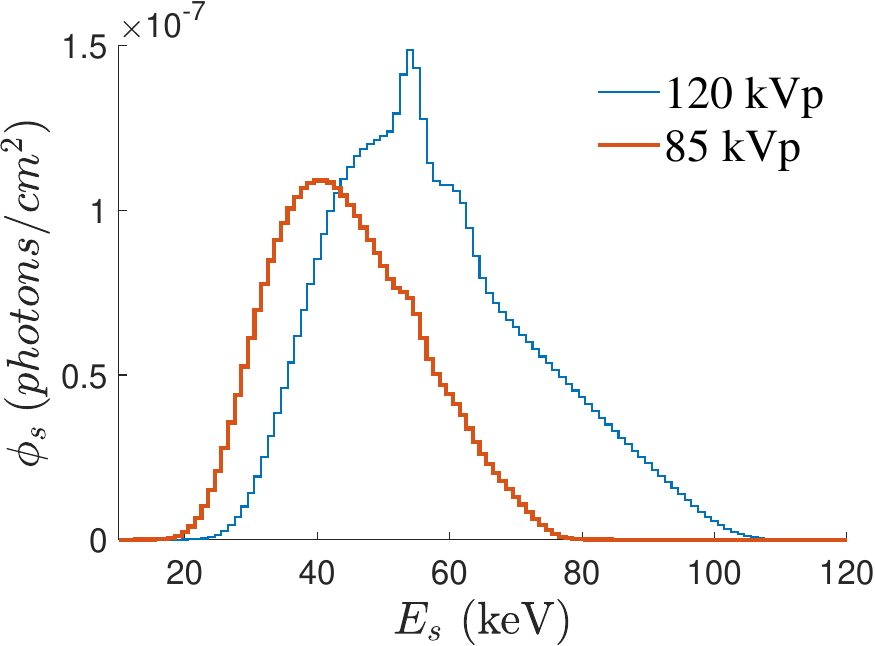}\label{fig:poly_out}}
\subfigure[LI at 1 m, 0.3 mm lead]
{\includegraphics[width=0.49\textwidth]{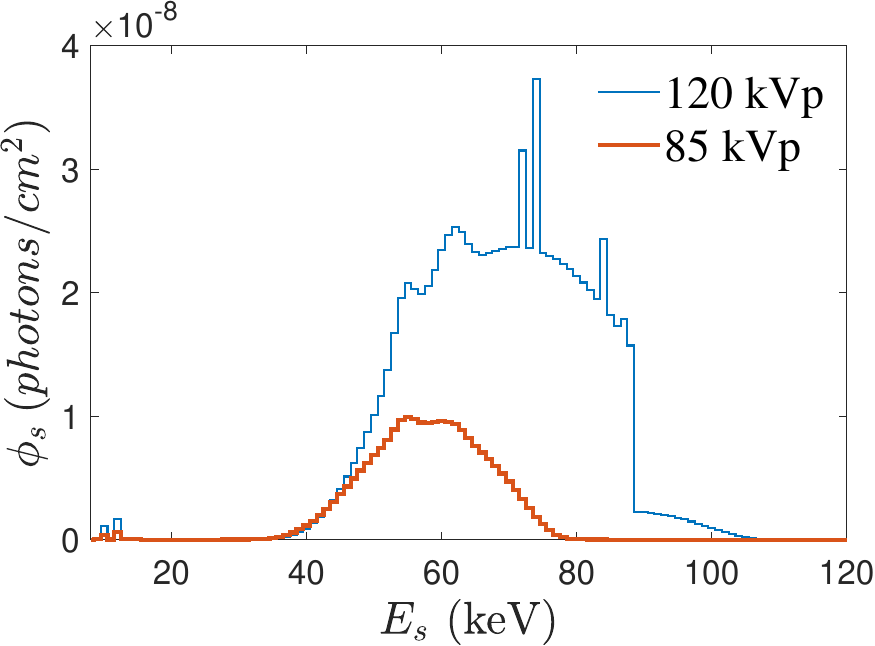}\label{fig:li}}
\subfigure[BS at 3 m]
{\includegraphics[width=0.49\textwidth]{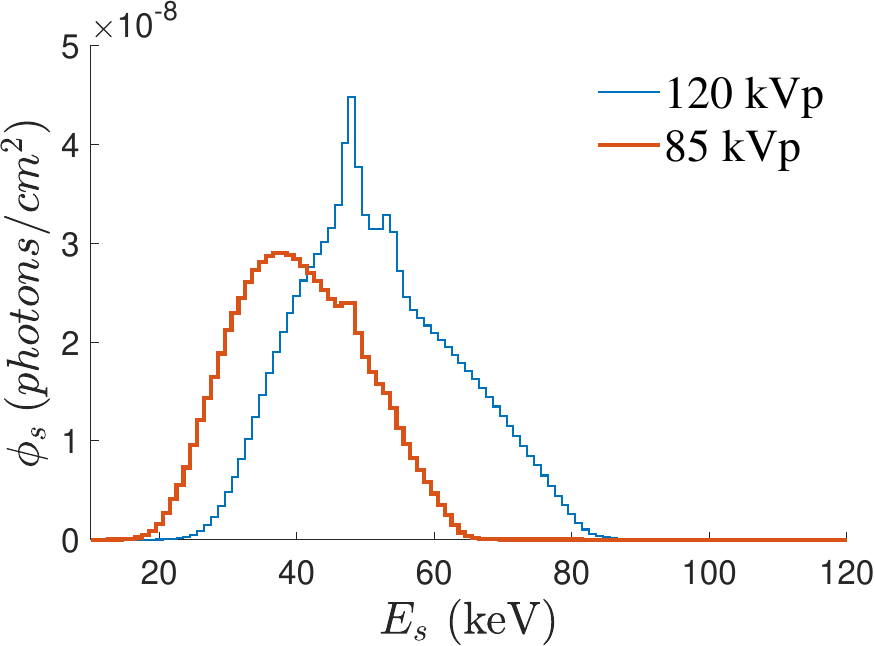}\label{fig:poly_BS}}
\subfigure[BS at 3 m, 15 cm concrete]
{\includegraphics[width=0.49\textwidth]{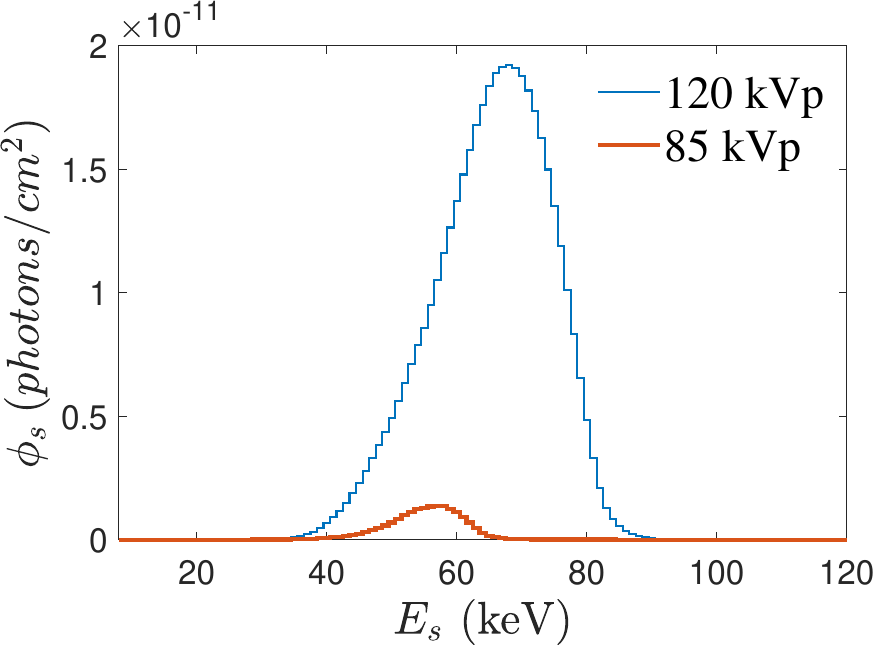}\label{fig:bs}}
\subfigure[FS at 3m]
{\includegraphics[width=0.49\textwidth]{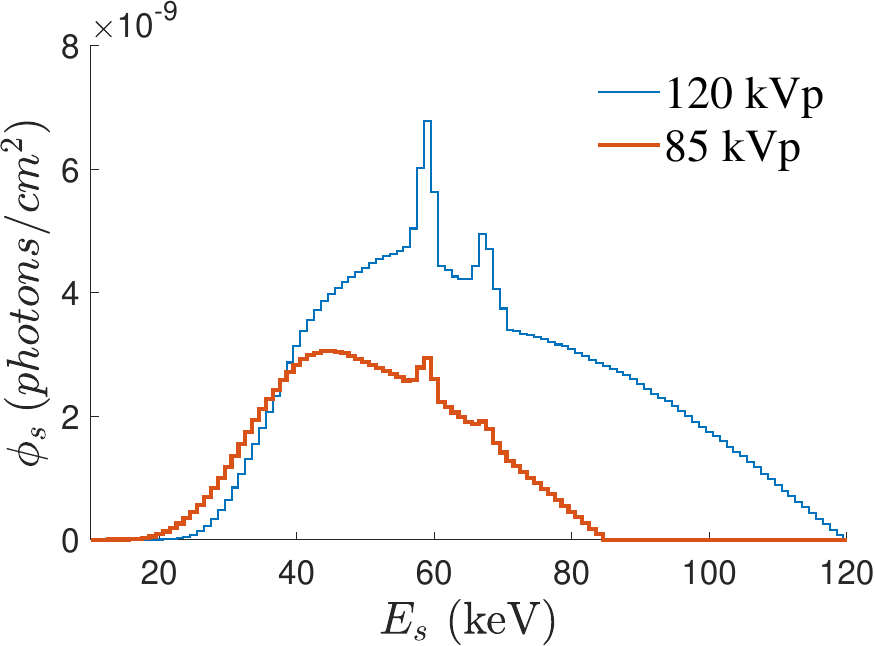}\label{fig:poly_FS}}
\subfigure[FS at 3m, 15 cm concrete]
{\includegraphics[width=0.49\textwidth]{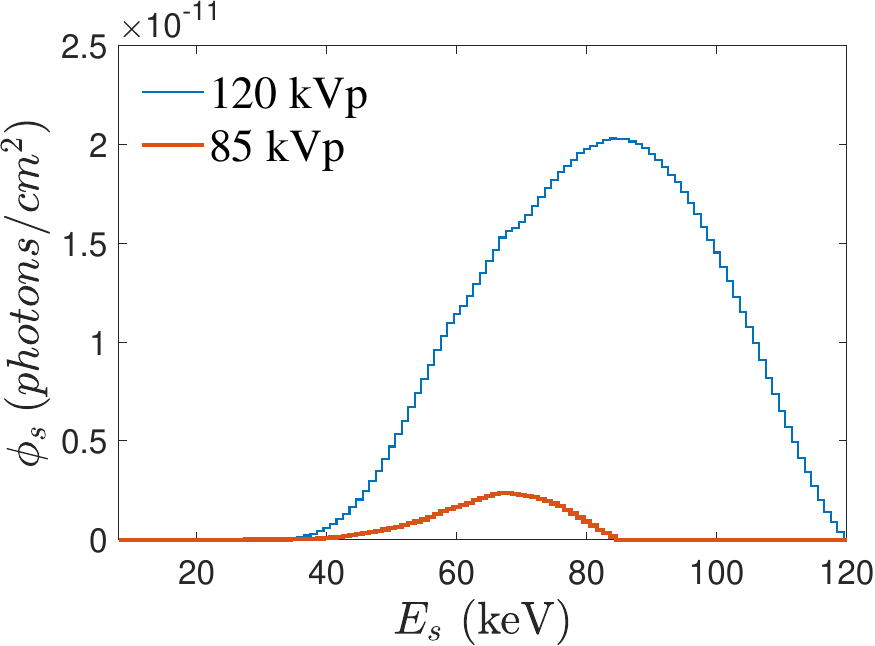}\label{fig:fs}}
\caption{Left: Fluence spectra emerging from a water phantom in different directions and at different distances from the phantom (per primary emitted photon). Right: Fluence spectra after being attenuated by 0.3 mm lead shielding and 15 cm concrete shielding. The directions are lateral-inferior (LI), back-scatter (BS), and forward-scatter (FS) \cite{GonzalezLopez2022}.}
\label{fig:phantom}
\end{figure}

Figure \ref{fig:phantom} shows, in its left column, spectral fluences generated in water phantoms \cite{GonzalezLopez2022} and directed towards personnel in an interventional radiology suite (figure \ref{fig:poly_out}), and towards the ceiling (figure \ref{fig:poly_BS}) and floor (figure \ref{fig:poly_FS}) of the room. In its right column, spectral fluences attenuated by lead thicknesses similar to those present in lead aprons \ref{fig:li} and concrete thicknesses similar to those found in ceilings \ref{fig:bs} and floors \ref{fig:fs} of the rooms are shown.

\begin{figure}
\centering
{\includegraphics[width=1\textwidth]{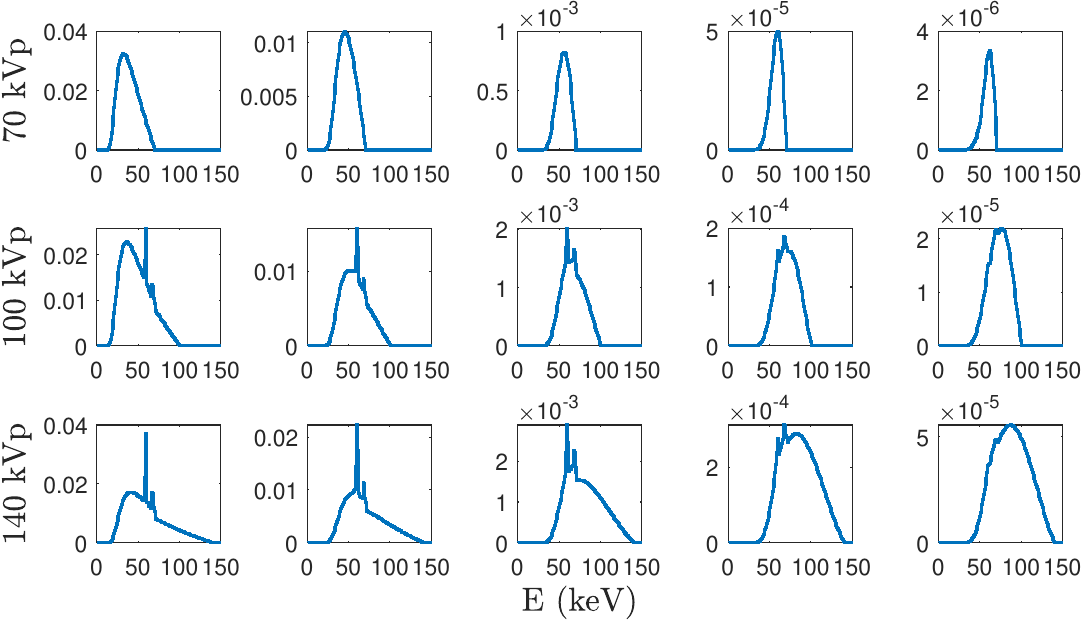}}
\caption{Energy spectra of incident beams on concrete barriers of different thicknesses (first column). Energy spectra of beams transmitted through 1, 5, 10 and 15 cm concrete barriers (second, third, fourth and fifth columns, respectively). All spectra are normalized to 1 photon incident on the barrier.}
\label{fig:espectros_hormigon}
\end{figure}

\begin{figure}
\centering
{\includegraphics[width=0.75\textwidth]{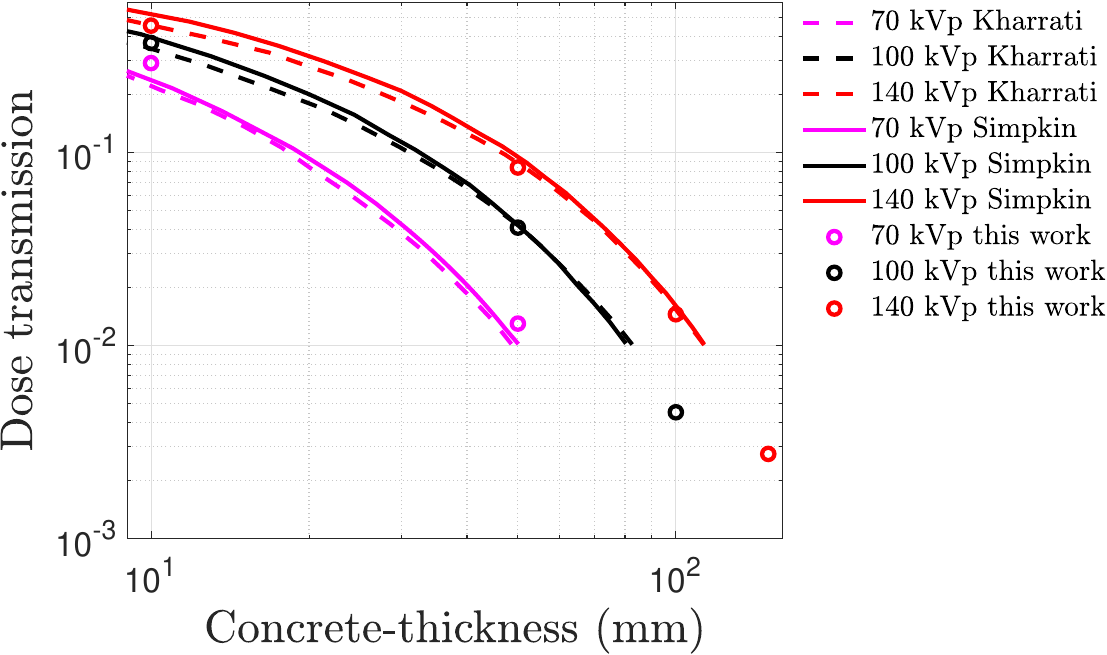}}
\caption{Comparison of dose transmission obtained in this work with the PENELOPE code \cite{PENELOPE} with results produced by other Monte Carlo codes. Kharrati et al. used Monte Carlo N-Particle (MCNP) \cite{mcnp} and Simpkin used Electron Gamma Shower (EGS4)\cite{egs4}.}
\label{fig:compara_hormigon}
\end{figure}

Figure \ref{fig:espectros_hormigon} shows how the energy spectra of incident beams transform as they pass through concrete barriers of 1, 5, 10 and 15 cm of thickness. These beams were used to compare dose transmission with previously published results. The comparison of the results obtained for the transmission of polyenergetic beams through concrete barriers is presented in Figure \ref{fig:compara_hormigon}. The values calculated in this work are shown alongside those obtained by Kharrati \cite{Kharrati2007} and Simpkin \cite{simpkin1989}.

\section{Discussion}
When a monoenergetic beam perpendicularly strikes a flat barrier, the spectrum of the emerging radiation varies depending on the energy of the incident photons, the material of the barrier, and its thickness. Figure \ref{fig:spectra} illustrates how differently the barrier can affect the spectrum. As seen in figure \ref{fig:t_0p15cmLead89kVp}, the secondary radiation produced by fluorescence largely dominates the primary radiation transmitted through the barrier, contrary to what happens in the rest of the cases shown. This is a clearly distinctive characteristic of the effect of lead barriers compared to those of concrete or other materials found in x-ray facilities: for incident photons of higher energy, the radiation transmitted through the barrier contains a large proportion of fluorescent photons with energies limited by the K-edge energy of lead.

It's worth noting the different decay observed in the radial coordinate between lead and concrete (figure \ref{fig:spectra}). For radii of a few tenths of millimeters the number of photons becomes negligible in the case of lead and incident energy of 60 keV. This range extends to several centimeters for the concrete case. Also, the number of secondary photons increases at higher $r$ values as the thickness of the barrier or the incident energy increases, consistent with findings from a previous study \cite{GonzalezLopez2023}.

By excluding lead from the analysis, the materials exhibit similar behavior (Figure \ref{fig:materiales}), as they do not contain elements that exhibit K-shell fluorescence within the presented energy range. On the other hand, the large proportion of calcium in gypsum increases the material's effective atomic number compared to concrete, making it more absorbent to radiation.

In a previous study \cite{GonzalezLopez2022}, secondary radiation spectra were determined in water phantoms simulating patients. The method was similar to that used in the present study, collecting photons scattered in different directions and grouping them according to their energy and direction. In this way, and using primary radiation beams with sizes and spectra typical of x-ray beams used in diagnosis, it was possible to characterize the beams reaching different areas of an x-ray room. Combining the results of that study with those of this work, the performance of protection barriers typically used in x-ray facilities can be calculated. In this regard, figure \ref{fig:phantom} illustrates the transformation undergone by the radiation spectrum after passing through shields of the type commonly used in practice. Therefore, it is possible to calculate the reduction in air kerma resulting from the use of these protection barriers.

As shown in Figure \ref{fig:compara_hormigon}, the calculation method presented in this work, based on monoenergetic pencil beams, accurately reproduces the dose transmission results obtained by Kharrati \cite{Kharrati2007} and Simpkin \cite{simpkin1989}.

\section{Conclusions}
In this work, Monte Carlo simulations have been employed to characterize the spectra of emerging photons from shielding barriers made of various materials and thicknesses. For this characterization, monoenergetic pencil photon beams with energies within the diagnostic imaging energy range have been utilized.

The composition of the materials present in a protective barrier against ionizing radiation in the energy range of diagnostic radiology determines the spectrum of the emerging radiation. The effect of the materials can result in greater or lesser attenuation of the emerging photons across the entire spectrum, as observed with glass, gypsum, clay, and concrete. Alternatively, it can radically transform the shape of the spectrum, as occurs with lead at energies slightly above the K-fluorescence threshold, causing the transmitted energy due to secondary radiation to become predominant.

The characterization of emerging spectra from monoenergetic pencil beams allows for calculations in a wide range of practical cases: On one hand, knowing the emerging spectra for monoenergetic incident beams allows for the calculation of the emerging spectrum of any polyenergetic incident beam using Equation \ref{eq:poly}. On the other hand, extending the results obtained for pencil beams to broad beams is a direct consequence of the linear and spatially invariant nature (in any plane parallel to the studied barriers) of the interaction described in the simulations.

\end{document}